\documentclass[12pt]{article}
\usepackage{amsfonts}
\usepackage{amsmath}
\usepackage{amssymb}
\usepackage{graphicx}
\usepackage{color}
\usepackage[all, knot]{xy}
\usepackage{tikz}
\usepackage{array}

\usepackage{ulem}

\usepackage[utf8]{inputenc}
\usepackage{epstopdf}
\usepackage[footnotesize]{caption}
\usepackage{amsthm}
\usepackage{enumitem}
\usepackage{mathrsfs}

\usepackage[margin=3cm]{geometry}

\def \be {\begin{equation}}
\def \ee {\end{equation}}
\def \bea {\begin{eqnarray}}
\def \eea {\end{eqnarray}}
\def \nn {\nonumber}

\def \rr {\raise.35ex\hbox{\small $\prime$}\kern-.17em{\mbox{\large $\imath$}}}

\def \dels {\partial\kern-.6em /\kern.1em}
\def \As {{A\kern-.5em / \kern.5em}}
\def \Ds {D\kern-.7em / \kern.5em}

\def \ks {k\kern-.5em /}
\def \ls {l\kern-.5em /}

\newcommand{\ci}[1]{}

\newcommand{\ba}{\begin{eqnarray}}
\newcommand{\ea}{\end{eqnarray}}
\newcommand{\bal}{\begin{align}}
\newcommand{\eal}{\end{align}}
\newcommand{\bay}[1]{\left(\begin{array}{#1}}
\newcommand{\eay}{\end{array}\right)}

\newcommand{\hide}[1]{}

\newlist{axioms}{enumerate}{2}
\setlist[axioms,1]{label=\textbf{A\arabic{axiomsi}.}, ref=A\arabic{axiomsi}}
\setlist[axioms,2]{label=\textbf{A\arabic{axiomsi}\rlap{\myEnumCounter{axiomsii}}.},%
                   ref=A\arabic{axiomsi}\myEnumCounter{axiomsii},%
                   align=parleft,%
                   leftmargin=0em,%
                   itemsep=1.4ex,%
                   before={\stepcounter{axiomsi}}}

  \usetikzlibrary{decorations.markings}

\begin{document}
\begin{titlepage}

\begin{center}

\textbf{\LARGE
Naive Lattice Fermion without Doublers
\vskip.3cm
}
\vskip .5in
{\large
Xingyu Guo$^{a, b}$ \footnote{e-mail address: guoxy@m.scnu.edu.cn}, 
Chen-Te Ma$^{a, b, c, d, e}$ \footnote{e-mail address: yefgst@gmail.com}, 
and Hui Zhang$^{a, b}$ \footnote{e-mail address: Mr.zhanghui@m.scnu.edu.cn}
\\
\vskip 1mm
}
{\sl
$^a$
Guangdong Provincial Key Laboratory of Nuclear Science,\\
 Institute of Quantum Matter,
South China Normal University, Guangzhou 510006, Guangdong, China.
\\
$^b$
Guangdong-Hong Kong Joint Laboratory of Quantum Matter,\\
 Southern Nuclear Science Computing Center, 
South China Normal University, Guangzhou 510006, Guangdong, China.
\\
$^c$
Asia Pacific Center for Theoretical Physics,\\
Pohang University of Science and Technology, 
Pohang 37673, Gyeongsangbuk-do, South Korea. 
\\
$^d$
School of Physics and Telecommunication Engineering,\\ 
South China Normal University, Guangzhou 510006, Guangdong, China.
\\
$^e$
The Laboratory for Quantum Gravity and Strings,\\
 Department of Mathematics and Applied Mathematics,\\
University of Cape Town, Private Bag, Rondebosch 7700, South Africa.
}
\\
\vskip 1mm
\vspace{40pt}
\end{center}
\newpage
\begin{abstract}
We discuss the naive lattice fermion without the issue of doublers. 
A local lattice massless fermion action with chiral symmetry and hermiticity cannot avoid the doubling problem from the Nielsen-Ninomiya theorem. 
Here we adopt the forward finite-difference deforming the $\gamma_5$-hermiticity but preserving the continuum chiral-symmetry. 
The lattice momentum is not hermitian without the continuum limit now. 
We demonstrate that there is no doubling issue from an exact solution. 
The propagator only has one pole in the first-order accuracy. 
Therefore, it is hard to know the avoiding due to the non-hermiticity. 
For the second-order, the lattice propagator has two poles as before. 
This case also does not suffer from the doubling problem. 
Hence separating the forward derivative from the backward one evades the doublers under the field theory limit. 
Simultaneously, it is equivalent to breaking the hermiticity.
In the end, we discuss the topological charge and also demonstrate the numerical implementation of the Hybrid Monte Carlo.   
\end{abstract}
\end{titlepage}

\section{Introduction}
\label{sec:1}
\noindent 
It is {\it hard} to have an analytical solution in strongly coupled systems. 
For studying physics, people adopted lattice regularization for putting the systems on a lattice. 
In the high-energy community, many fundamental problems rely on a study of the Quantum Chromodynamics (QCD) model. 
However, due to a lack of analytical tools, people need to rely on a lattice method for exploring the QCD model. 
\\

\noindent 
The {\it gauge sector} of Lattice Quantum Chromodynamics (LQCD) keeps the gauge invariance for any {\it finite} lattice spacing ($a$) \cite{Wilson:1974sk}. 
The lattice gauge symmetry leads to a practical simulation method. 
However, LQCD is still time-consuming due to the {\it Nielsen-Ninomiya theorem} \cite{Nielsen:1980rz, Nielsen:1981xu, Karsten:1981gd}. 
This theorem prohibits a {\it hermitian} construction of $d$-dimensional fermion lattice action $S_F$: 
\begin{itemize}
\item{1. $D(x)$ is exponentially local, which implies that the operator is bounded by $\sim\exp(-|x|/c)$, where $c\propto a$ ;
}
\item{2. $\bar{D}(p)=i\gamma_{\mu}p_{\mu}+{\cal O}(ap^2)$ for $p\ll\pi/a$;
}
\item{3. $\bar{D}(p)$ is invertible for $p\neq 0$ (no massless doublers);
}
\item{4. $\gamma_5D+D\gamma_5=0$ (continuum chiral symmetry), 
}
\end{itemize}
where $D(x)$ is a Dirac matrix satisfying 
\bea
S_F=a^d\sum_{\mathrm{all\ lattice\ points}}\bar{\psi}(D+m)\psi.
\eea
The $\bar{D}(p)$ is a Dirac matrix on a momentum space, the $\bar{\psi}$ and $\psi$ are independent Dirac fermion fields, and $\gamma_{\mu}$ and $\gamma_5$ are the gamma matrices.  
The $m$, $x$, and $p_{\mu}$ are the fermion mass, position, and momenta, respectively. 
We label the spacetime indices by $\mu=1, 2, \cdots, d$.  
A Dirac fermion lattice theory should require the first {\it three} conditions. 
People could {\it avoid} the no-go by modifying the {\it chiral symmetry} condition (like overlap fermion). 
However, it is necessary to accept the {\it square root} operation with a lattice chiral-symmetry. 
Another way is to introduce the {\it non-physical} degrees of freedom but lose chiral symmetry (like Wilson-Dirac fermion). 
It should be problematic for studying a {\it light} fermion mass. 
Hence the {\it no-go} provides a {\it strong constraint} to the construction of lattice fermion. 
The {\it square-root} operation or {\it losing} chiral symmetry all introduce practical problems about {\it simulation time} or {\it error bar}.
\\

\noindent
In Refs. \cite{Stamatescu:1993ga, Stamatescu:1994yj}, one used a one-sided lattice difference with the first-order accuracy to show the naive lattice fermion with a chiral symmetry. 
In this explicit example, the lattice momentum operator is not hermitian, except for the continuum limit \cite{Stamatescu:1993ga}. 
The lattice action loses the hermiticity without violating the no-go. 
Hence breaking the hermicity seems to be the successful reason for solving the fermion doubling problem. 
However, the first-order accuracy only provides one pole to the propagator. 
It is also hard to argue that one pole is due to the non-hermiticity. 
The one-sided lattice difference breaks the hypercubic symmetry \cite{Stamatescu:1993ga}. 
Therefore, the lattice interacting field theory suffers the issue of non-renormalizability \cite{Sadooghi:1996ip}. 
Hence it is necessary to impose the averaging over all possible one-sided derivatives \cite{Stamatescu:1993ga} to remove the non-renormalizable terms \cite{Sadooghi:1996ip}.
\\

\noindent
The central question that we would like to address in this letter is the following: {\it How to construct and implement naive lattice fermions without the doubling problem?} 
For a {\it 1d lattice fermion system}, we can calculate all integration exactly for each {\it finite} size.  
Hence we can explicitly study the doubling problem of the {\it first-order accuracy} and the {\it second-order accuracy} for the {\it forward finite-difference}. 
Both cases lose {\it hermiticity} on a {\it lattice}. 
The second-order provides {\it two} poles to the propagator. 
Hence it should be the best way to justify the non-hermiticity. 
Ones also proposed that using the {\it bi-orthogonal basis} realizes the {\it index theorem} \cite{Atiyah:1968mp, Atiyah:1970ws, Atiyah:1971rm} on a lattice \cite{Xi:2020iji}. 
The index theorem helps extract the zero-modes of a Dirac matrix to obtain a {\it topological charge}.  
One already showed that a consistent construction of the Dirac matrix does {\it not} necessarily generate a correct topological charge on a lattice \cite{Chiu:2001bg}. 
The {\it exponentially-local}, {\it doublers-free}, and a {\it correct continuum behavior} in the Dirac matrix should just guarantee a correct {\it homogeneous-solution} of topological charge density under the field theory limit (infinite size and continuum limits) \cite{Chiu:2001ja}. 
For studying non-perturbative physics in the QCD model, non-trivial topological charges should {\it not} lose \cite{Fujikawa:1979ay}. 
Hence analyzing the definition of topological charge is necessary for a numerical implementation \cite{Chiu:1998bh}. 
\\

\noindent
In this letter, we show that the {\it second-order accuracy} evades the {\it fermion doubling problem} without breaking the chiral symmetry. 
From the study of the exact solution, we understand that the doubling problem occurs due to a combination of {\it forward} and {\it backward} finite-difference. 
After we only adopt {\it one} finite-difference scheme, it is equivalent to {\it decoupling} the non-physical poles from the physical one (under the field theory limit). 
This approach also directly brings a {\it broken} of the hermiticity.
Therefore, we conclude that breaking the hermiticity should be an elegant idea for escaping the doubling problem. 
We also show that the {\it lattice index theorem} only brings the {\it trivial} topological charge. 
In the end, we demonstrate a numerical implementation by the {\it Hybrid Monte Carlo} for two lattice fermions with a degenerate mass in 1d. 

\section{1d Lattice Fermion}
\label{sec:2}
\noindent 
We first review the continuum theory. 
We then show exact solutions for forward finite-difference in 1d lattice fermion. 
The result is consistent with the continuum physics avoiding the fermion doubling problem.

\subsection{Continuum Theory}
\noindent
We first introduce the 1d Dirac fermion continuum theory. 
The Euclidean action is $S_{FC}=\int dx\ \bar{\psi}(x)(\gamma_1\partial_1+m)\psi(x)$,
where
\bea
\gamma_1\equiv
\begin{pmatrix}
1&0
\\
0&-1
\end{pmatrix}; \qquad 
\partial_1\equiv\frac{\partial}{\partial x}. 
\eea 
The propagator satisfies $\big(\gamma_1(d/dx)+m\big)S(x)=\delta(x)$.
The solution is:
\bea
S(x)=\int_{-\infty}^{\infty}\frac{dp}{2\pi}\ e^{ipx}\frac{1}{i\gamma_1p+m}
=
\begin{pmatrix}
\theta(x)&0
\\
0&\theta(-x)
\end{pmatrix}
e^{-m|x|},
\eea
where
\bea
\theta(x)\equiv\Bigg\{\begin{array}{ll}
                 1, & x\ge 0 \\  
                 0, & x<0
                \end{array}.
\eea
The fermion doubling problem generates one non-physical pole in the 1d lattice fermion. 
This pole gives a non-vanishing contribution even under the field theory limit.  
Therefore, we cannot obtain a correct continuum limit. 
Later we will compare the result of a forward one to the continuum result.  

\subsection{1st Order} 
\noindent 
Now we adopt the forward finite-difference with the first-order accuracy to write the following naive lattice action: 
\bea
S_{F1}&=&a\sum_{n=0}^{N-1}\bar{\psi}(n)\bigg(\gamma_1\frac{\psi(n+1)-\psi(n)}{a}+m\psi(n)\bigg)
\nn\\
&=&a\sum_{n_1, n_2; \alpha_1, \alpha_2}\bar{\psi}(n_1)_{\alpha_1}\big(D(n_1, n_2)_{\alpha_1, \alpha_2}+m\delta_{n_1, n_2}\delta_{\alpha_1, \alpha_2}\big)\psi(n_2)_{\alpha_2},
\eea
where $D(n_1, n_2)_{\alpha_1, \alpha_2}\equiv
(\gamma_1)_{\alpha_1, \alpha_2}(\delta_{n_1+1, n_2}-\delta_{n_1, n_2})/a.$
We label the matrix components of $\gamma_1$ by $\alpha_1, \alpha_2=1, 2$. 
The $N$ is the number of lattice points. 
The lattice fermion field satisfies the anti-periodic boundary condition $\psi(0)=-\psi(N)$.
\\

\begin{figure}
\includegraphics[width=1.\textwidth]{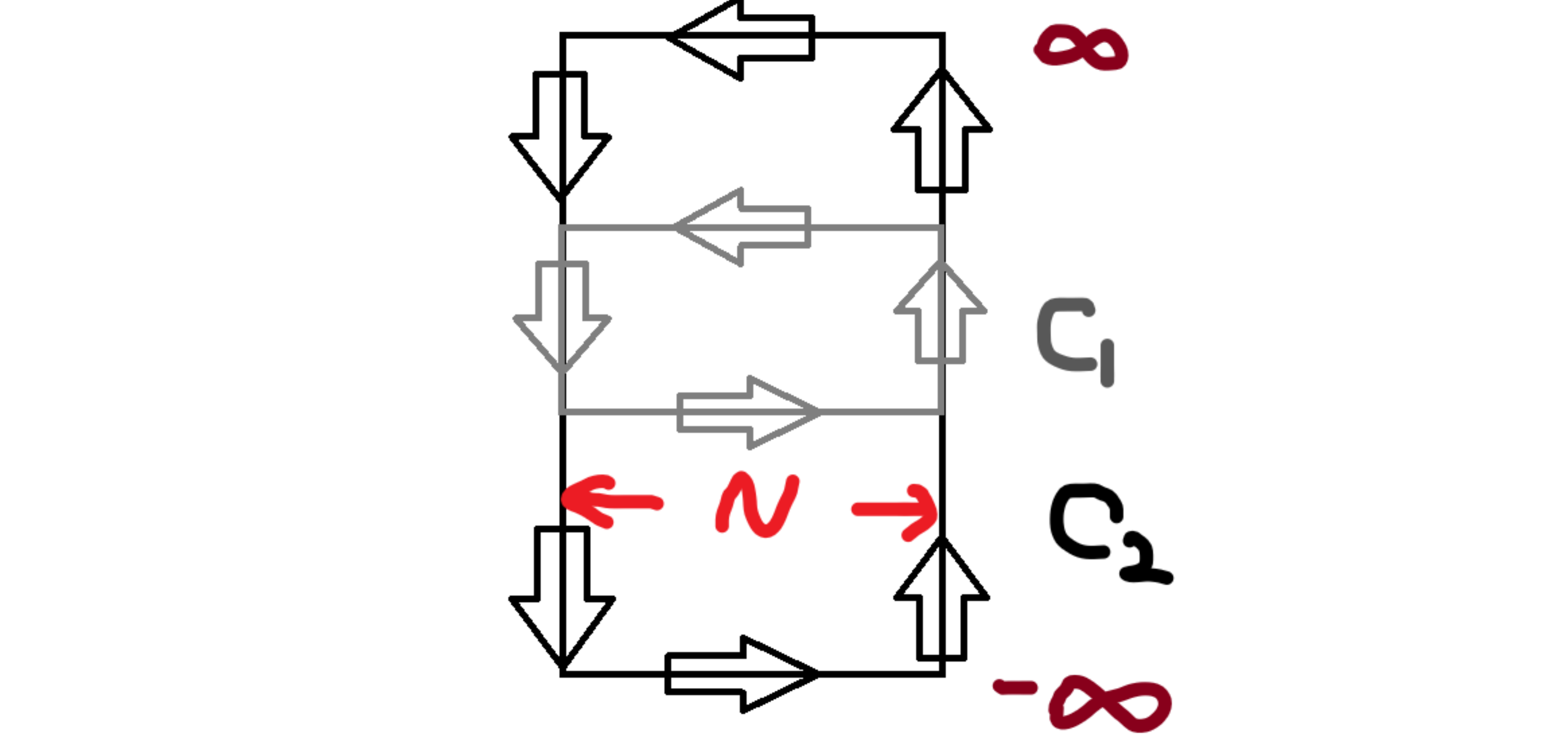}
\caption{We show the complex contours $C_1$ and $C_2$. 
The length of horizon direction for each contour is $N$, where $N$ is the number of lattice points. 
The boundary of vertical direction for the $C_2$ tends to $\infty$ and $-\infty$. 
\label{CC}}
\end{figure}

\noindent
We first show the lattice propagator for $x\equiv na> 0$ as the following: 
\bea
&&
S_{L}(x)
\nn\\
&=&\frac{1}{N}\sum_{j=0}^{N-1}\exp\bigg(i\frac{(2j+1)\pi}{N}n\bigg)
\begin{pmatrix}
\frac{1}{\exp\big(i\frac{(2j+1)\pi}{N}\big)-1+ma}&0
\\
0&\frac{1}{-\exp\big(i\frac{(2j+1)\pi}{N}\big)+1+ma}
\end{pmatrix}
\nn\\
&=&
\frac{1}{N}\oint_{C_1} dw\ \exp\bigg(i\frac{2w\pi}{N}n\bigg)
\begin{pmatrix}
\frac{\frac{-1}{\exp(2\pi i w)+1}}{\exp\big(i\frac{2w\pi}{N}\big)-1+ma}&0
\\
0&\frac{\frac{-1}{\exp(2\pi i w)+1}}{-\exp\big(i\frac{2w\pi}{N}\big)+1+ma}
\end{pmatrix}
\nn\\
&=&
\begin{pmatrix}
\frac{(1-ma)^{n-1}}{(1-ma)^N+1}& 0
\\
0& -\frac{(1+ma)^{n-1}}{(1+ma)^{N}+1}
\end{pmatrix}
\nn\\
&\equiv& 
\begin{pmatrix}
S_{11}(x)& 0
\\
0& S_{22}(x)
\end{pmatrix}.
\eea 
The closed loop $C_1$ encloses the poles $w=1/2, 3/2 \cdots, (N-1/2)$. 
For $x<0$, we need to replace $\exp(2\pi i w)$  with $\exp(-2\pi i w)$ without a divergent boundary. 
The lattice propagator becomes: 
\bea
S_{L}(x)
&=&
\frac{1}{N}\oint_{C_1} dw\ \exp\bigg(i\frac{2w\pi}{N}n\bigg)
\begin{pmatrix}
\frac{\frac{1}{\exp(-2\pi i w)+1}}{\exp\big(i\frac{2w\pi}{N}\big)-1+ma}&0
\\
0&\frac{\frac{1}{\exp(-2\pi i w)+1}}{-\exp\big(i\frac{2w\pi}{N}\big)+1+ma}
\end{pmatrix}
\nn\\
&=&
\begin{pmatrix}
-\frac{(1-ma)^{n-1}}{(1-ma)^{-N}+1}& 0
\\
0& \frac{(1+ma)^{n-1}}{(1+ma)^{-N}+1}
\end{pmatrix}.
\eea 
We include all poles by the contour $C_2$ (as shown in Fig. \ref{CC}). 
Because the contour integration of $S_L$ is invariant under $w\rightarrow w+N$, the contour integration alone the $C_2$ vanishes. 
In other words, we use another pole to calculate the contour integration alone $C_1$.  
For analyzing the number of poles, we first take the infinite size limit ($N\rightarrow\infty$). 
Here we are interested in the continuum result. 
Therefore, we only consider $ma<1$ in this letter. 
The lattice propagator in the infinite size limit is:
\bea
S_{11}(x)&\rightarrow&\Bigg\{\begin{array}{ll}
                 (1-ma)^{n-1}, & x> 0 \\  
                 0, & x<0
                \end{array};
\nn\\
S_{22}(x)&\rightarrow&\Bigg\{\begin{array}{ll}
                 0, & x> 0 \\  
                 (1+ma)^{n-1}, & x<0
                \end{array}.
\eea
Here we show that the non-physical contribution vanishes under the infinite lattice size limit.
We then take the continuum limit ($ma, a/x\rightarrow 0$). 
The lattice propagator is the same as the $S(x)$. 
From the study, we now only have one pole. 
Because we replace $i\sin(pa)$ with $\exp(ipa),$ the number of poles reduces by half compared to the doubling case. 
The sine function appeared before due to a combination of the forward and backward finite-difference. 
Each dimension has only one pole now. 
Therefore, considering the general $d$-dimension also evades the doubling problem.
\\

\noindent 
It is easy to show that the Dirac matrix of the naive lattice fermion satisfies the chiral symmetry condition 
\bea
\gamma_5D+D\gamma_5=0.
\eea 
In 1d case, we choose 
\bea
\gamma_5\equiv\begin{pmatrix}
0& 1
\\
1& 0
\end{pmatrix}. 
\eea 
It is necessary to apply a non-symmetrized way to define a lattice derivative for a naive fermion.
Indeed, it leads to a breakdown of the $\gamma_5$-hermiticity. 
We demonstrate this fact explicitly in 1d case:
\bea
\gamma_5D\gamma_5=-D; \qquad D^{\dagger}-\gamma_5D\gamma_5=D+D^{\dagger}. 
\eea 
Because the Dirac matrix is at the order of $1/a$, we cannot take the continuum limit to recover the $\gamma_5$-hermicity in general. 
Now we show that the $\gamma_5$-hermicity can recover for the physical modes of the fermion field (the eigenvalues of $D$ are finite under the continuum limit). 
To give general proof, we first introduce the gauge field as in the following:
\bea
S_{LG}=a\sum_{n=0}^{N-1}\bar{\psi}(n)\bigg(\gamma_{1}\frac{U_{1}(n)\psi(n+1)-\psi(n)}{a}+m\psi(n)\bigg). 
\eea
The Dirac matrix becomes $D(j, k)=\gamma_{1}\big(U_{1}(n)\delta_{j+1, k}-\delta_{j, k}\big)/a$.
The gauge link is \cite{Wilson:1974sk}
\bea
U_{1}(n)\equiv e^{iaA_{1}(n)},
\eea  
where $A_1$ is the gauge field. 
We show that 
\bea
\big(D+D^{\dagger}\big)\psi={\cal O}(a)
\eea
from the following expansion:
\bea
U_1(j\pm 1)=1+iaA_1(j)+{\cal O}(a^2); \qquad \psi(j\pm 1)=\psi(j)\pm a\psi^{\prime}(j)+{\cal O}(a^2), 
\eea
where $\psi^{\prime}$ is the derivative of a fermion field. 
\\

\noindent
For evading the no-go, people used a Ginsparg-Wilson relation defining a lattice chiral-symmetry.
Under the continuum limit, the Ginsparg-Wilson relation reducing to chiral symmetry is only for physical modes. 
Now we use the Wilson-Dirac fermion to demonstrate the problem of non-physical mode
\bea
S_{WD}=a\sum_{n=0}^{N-1}\bar{\psi}(n)
\big(\gamma_1\otimes(\tilde{D}-\tilde{D}^{\dagger})+I\otimes(\tilde{D}+\tilde{D}^{\dagger}+m)\big), 
\eea 
where $\tilde{D}$ is the forward finite-difference of $\partial_1$ at the first order. 
The backward finite-difference of $\partial_1$ with the same accuracy is equivalent to $-\tilde{D}^{\dagger}$. 
The $\tilde{D}+\tilde{D}^{\dagger}$ is the familiar Wilson mass term. 
Therefore, we can find that the the first diagonal block of the Dirac matrix corresponds to the forward finite difference. 
The second diagonal block corresponds to the backward finite difference. 
We know that the Wilson term introduces a mass to the non-physical mode, which has a non-vanishing contribution of the Wilson term. 
This mode does not vanish under the continuum limit but will decouple due to an infinite mass. 
Because the Wilson term is at the same order of $a$ with the chiral symmetry condition, a non-physical mode does not respect the chiral symmetry. 
Here the difference of forward and backward finite-difference provides the Wilson mass term. 
If one only adopts one finite-difference scheme, the non-physical mass goes away. 
The Dirac matrix has a manifest hermiticity but violates the chiral symmetry condition. 
Here we choose to modify $\gamma_5$-hermiticity. 
Preserving the chiral symmetry should be an advantage point compared to the Wilson-Dirac formulation. 
Realizing the chiral symmetry without a square root should reduce the simulation time.  
\\

\noindent
One cannot apply the Nielsen-Ninomiya theorem \cite{Nielsen:1980rz, Nielsen:1981xu, Karsten:1981gd} to this approach due to the non-hermiticity. 
However, it is hard to connect the non-hermiticity to the number of poles.  
In the second-order, the propagator has two poles. 
The lattice theory still loses hermiticity. 
We will show that only a physical pole survives under the field theory limit. 
It is one non-trivial example for showing that the non-hermiticity avoids the fermion doubling problem. 

\subsection{2nd Order}
\noindent
Using the second-order formula shows the lattice action
\bea
S_{F2}=\sum_{n=0}^{N-1}\bar{\psi}(n)\bigg\lbrack\gamma_1\bigg(-\frac{1}{2}\psi(n+2)+2\psi(n+1)-\frac{3}{2}\psi(n)\bigg)+m\psi(n)\bigg\rbrack.
\eea
For $x>0$, the lattice propagator is:
\bea
&&
S_{L}(x)
\nn\\
&=&\frac{1}{N}\sum_{j=0}^{N-1}\exp\bigg(i\frac{(2j+1)\pi}{N}n\bigg)
\nn\\
&&
\times
\begin{pmatrix}
\frac{1}{-\frac{1}{2}\exp\big(2i\frac{(2j+1)\pi}{N}\big)
+2\exp\big(i\frac{(2j+1)\pi}{N}\big)
-\frac{3}{2}+ma}&0
\\
0&\frac{1}{\frac{1}{2}\exp\big(2i\frac{(2j+1)\pi}{N}\big)
-2\exp\big(i\frac{(2j+1)\pi}{N}\big)
+\frac{3}{2}+ma}
\end{pmatrix}
\nn\\
&=&
\frac{1}{N}\oint_{C_1} dw\ \exp\bigg(i\frac{2w\pi}{N}n\bigg)
\nn\\
&&\times
\begin{pmatrix}
\frac{\frac{-1}{\exp(2\pi i w)+1}}{-\frac{1}{2}\exp\big(2i\frac{2w\pi}{N}\big)
+2\exp\big(i\frac{2w\pi}{N}\big)
-\frac{3}{2}+ma}&0
\\
0&\frac{\frac{-1}{\exp(2\pi i w)+1}}{\frac{1}{2}\exp\big(2i\frac{2w\pi}{N}\big)
-2\exp\big(i\frac{2w\pi}{N}\big)
+\frac{3}{2}+ma}
\end{pmatrix}
\nn\\
&=&
\begin{pmatrix}
F_1(n, m, N)-F_2(n, m, N)&0
\\
0&F_2(n, -m, N)-F_1(n, -m, N)
\end{pmatrix}, 
\eea
where 
\bea
F_1(n, m, N)&\equiv&\frac{(2-\sqrt{1+2ma})^n}{(2-\sqrt{1+2ma})^N+1}\frac{1}{(2-\sqrt{1+2ma})\sqrt{1+2ma}}; 
\nn\\
F_2(n, m, N)&\equiv&\frac{(2+\sqrt{1+2ma})^n}{(2+\sqrt{1+2ma})^N+1}\frac{1}{(2+\sqrt{1+2ma})\sqrt{1+2ma}}.
\eea
For $x<0$, the lattice propagator is:
\bea
&&
S_{L}(x)
\nn\\
&=&
\frac{1}{N}\oint_{C_1} dw\ \exp\bigg(i\frac{2w\pi}{N}n\bigg)
\nn\\
&&\times
\begin{pmatrix}
\frac{\frac{1}{\exp(-2\pi i w)+1}}{-\frac{1}{2}\exp\big(2i\frac{2w\pi}{N}\big)
+2\exp\big(i\frac{2w\pi}{N}\big)
-\frac{3}{2}+ma}&0
\\
0&\frac{\frac{1}{\exp(-2\pi i w)+1}}{\frac{1}{2}\exp\big(2i\frac{2w\pi}{N}\big)
-2\exp\big(i\frac{2w\pi}{N}\big)
+\frac{3}{2}+ma}
\end{pmatrix}
\nn\\
&=&
\begin{pmatrix}
-F_1(n, m, -N)+F_2(n, m, -N)&0
\\
0&-F_2(n, -m, -N)+F_1(n, -m, -N)
\end{pmatrix}. 
\nn\\
\eea
We first take the infinite lattice size limit for observing the contribution of poles. 
The lattice propagator becomes: 
\bea
S_{11}(x)&\rightarrow&\Bigg\{\begin{array}{ll}
                 (\frac{2-\sqrt{1+2ma})^{n-1}}{\sqrt{1+2ma}}, & x>0 \\  
                 \frac{(2+\sqrt{1+2ma})^{n-1}}{\sqrt{1+2ma}}, & x<0
                \end{array};
                \nn\\
S_{22}(x)&\rightarrow&\Bigg\{\begin{array}{ll}
                 0, & x>0 \\  
                 \frac{(2-\sqrt{1-2ma})^{n-1}}{\sqrt{1-2ma}}-\frac{(2+\sqrt{1-2m})^{n-1}}{\sqrt{1-2m}}, & x<0
                \end{array}.
\eea
The non-physical poles contribute to the lattice propagators as before. 
We then take the continuum limit, and the lattice propagator becomes the $S(x)$ as in the first-order case. 
The non-physical modes vanish only under the field theory limit. 
For a generalization of the higher dimensions, the conclusion is the same as the first-order accuracy. 
Hence we expect that breaking the hermiticity or $\gamma_5$-hermiticity should be proper for evading the fermion doubling problem. 
Our study also shows that one can use a higher-order accuracy to decrease the lattice artifact. 

\section{Topological Charge}
\label{sec:3}
\noindent
The motivation of lattice formulation is due to an interest in non-perturbative physics. 
Using the zero-mode of Dirac matrix does not guarantee to generate a correct topological charge \cite{Chiu:2001bg}. 
Therefore, it is necessary to discuss the topological charge before implementing the simulation. 
Here we show that using the bi-orthogonal basis \cite{Xi:2020iji} cannot give any non-trivial topological charge. 
\\

\noindent
The bi-orthogonal basis satisfies the following relations \cite{Xi:2020iji}:
\bea
\sum_{x, \alpha}\big(\phi_{L, j_1}^{\alpha}(x)\big)^*\phi_{R, j_2}^{\alpha}(x)&=&\delta_{j_1, j_2}; 
\nn\\ 
\sum_{j_1}\big(\phi_{L, j_1}^{\alpha}(x)\big)^*\phi_{R, j_1}^{\beta}(y)&=&\delta^{\alpha\beta}\delta_{x, y}; 
\nn\\
\sum_{y, \beta}D^{\alpha\beta}(x, y)\phi^{\beta}_{R, j_1}(y)&\equiv&\lambda_{j_1}\phi_{R, j_1}^{\alpha}(x); 
\nn\\ 
\sum_{y, \beta} \big(\phi^{\beta}_{L, j_1}(y)\big)^*D^{\beta\alpha}(y, x)&=&\lambda_{j_1}\big(\phi^{\alpha}_{L, j_1}(x)\big)^*, 
\nn\\
\eea
where $\phi_{L (R)}$ is the left (right)-eigenstate of $D$.
Ones can show that only zero-mode has a non-vanishing contribution for ${}_{L_j}\langle\gamma_5\rangle_{R_j}$ \cite{Xi:2020iji}: 
\bea
&&
\sum_{x, y, \alpha, \beta}\big(\phi^{\alpha}_{L, j}(x)\big)^*
(\gamma_5D)^{\alpha\beta}(x, y)
\phi^{\beta}_{R, j}(y)
\nn\\
&=&-\sum_{x, y, \alpha, \beta}\big(\phi^{\alpha}_{L, j}(x)\big)^*
(D\gamma_5)^{\alpha\beta}(x, y)
\phi^{\beta}_{R, j}(y); 
\nn\\
&&
\lambda_j{}_{L_j}\langle\gamma_5\rangle_{R_j}
\nn\\
&\equiv&\lambda_j\sum_{x, \alpha, \beta}\big(\phi^{\alpha}_{L, j}(x)\big)^*
(\gamma_5)^{\alpha\beta}
\phi^{\beta}_{R, j}(x)
\nn\\
&=&-\lambda_j\sum_{x, \alpha, \beta}\big(\phi^{\alpha}_{L, j}(x)\big)^*
(\gamma_5)^{\alpha\beta}
\phi^{\beta}_{R, j}(x).
\nn\\
\eea 
We then show that:
\bea
\sum_{x, j, \alpha, \beta}\big(\phi_{L, j}^{\alpha}(x)\big)^*(\gamma_5)^{\alpha\beta}\phi_{R, j}^{\beta}(x)
=\sum_{x, \alpha, \beta} (\gamma_5)^{\alpha\beta}\delta_{\alpha, \beta}
=0.  
\eea 
Therefore, we obtain the chirality sum rule $n_++N_+=n_-+N_-$, 
where $n_+$ ($n_-$) is the number of zero-mode of positive (negative) chirality, and 
$N_{\pm}$ denotes the non-zero mode case. 
Because we only have the zero-mode contribution, the lattice topological charge vanishes: 
\bea
Q_L\equiv n_-- n_+=0. 
\eea 
Our proof only requires that a Dirac matrix satisfies the continuum chiral symmetry condition. 
Therefore, the result also holds when considering all possible forward and backward derivatives in an interacting theory \cite{Stamatescu:1993ga, Sadooghi:1996ip}.  
\\

\begin{figure}[!htb]
\includegraphics[width=0.5\textwidth]{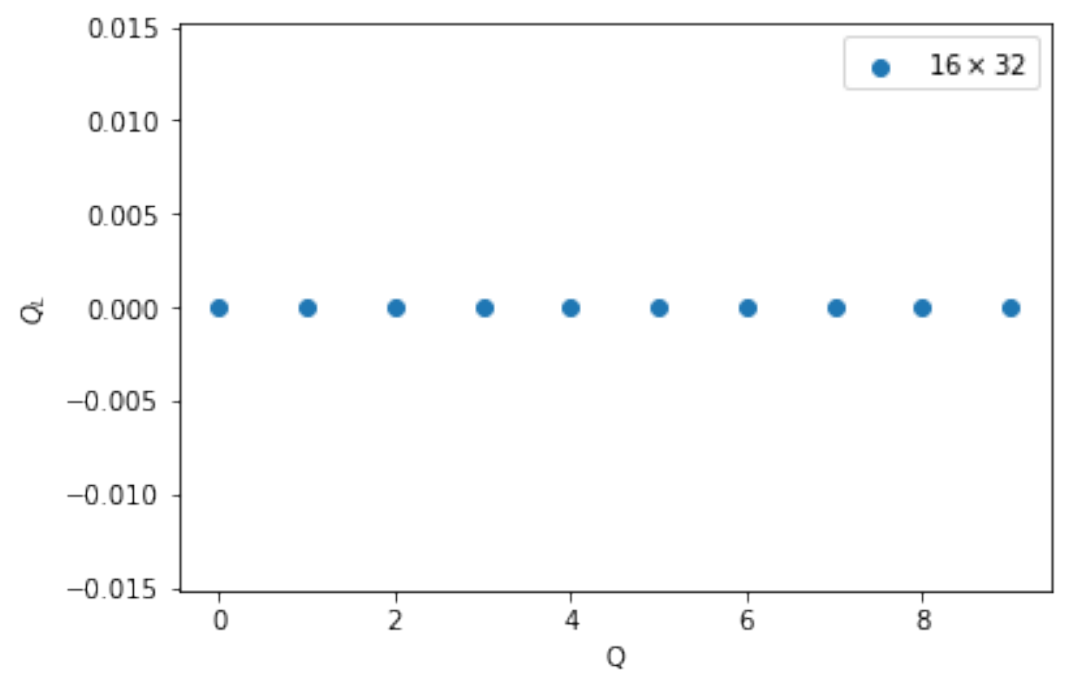}
\includegraphics[width=0.5\textwidth]{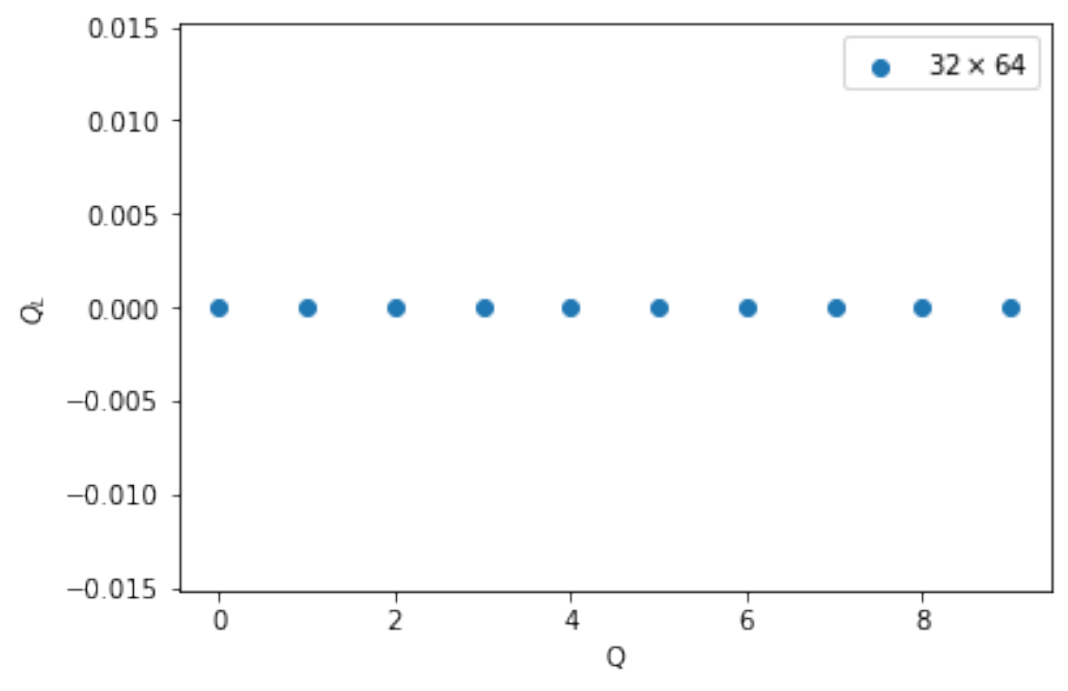}
\caption{We show the inconsistency between $Q_L$ and $Q$ for $(L_1, L_2)=(32, 16); (64, 32)$.} 
\label{1632}
\end{figure}

\noindent
When one applies Fujikawa's method \cite{Fujikawa:1979ay} to investigate the measure, the lattice measure is invariant under a chiral transformation. 
Indeed, obtaining the chiral anomaly is necessary to deform the chiral symmetry. 
The Ginsparg-Wilson relation introduces a topological charge on a chiral transformation. 
Therefore, Fujikawa's method on a lattice model will generate a chiral anomaly or topological charge term. 
The generation of non-trivial topological charge is also due to the non-vanishing contribution of ${}_{L_j}\langle\gamma_5\rangle_{R_j}$ from non-physical modes.
We confirm this proof from the following gauge configuration \cite{Chiu:1998bh} in Fig. \ref{1632}.
The gauge configuration is:
\bea
A_1(x)=-\frac{2\pi Q x_2}{L_1L_2}; \qquad 
A_2(x)=0, 
\eea
where $Q$ is the topological charge, $L_{\mu}\equiv N_{\mu}a$, and $x_{\mu}=0, a, \cdots, (N_{\mu}-1)a$. 
Here we choose the gauge links: 
\bea
U_{E, 1}(x)&\equiv& \exp\big(iA_1(x)a\big); 
\nn\\ 
U_{E, 2}(x)&\equiv& \exp\bigg(iA_2(x)a+i\frac{2\pi Q x_1}{L_1}\delta_{x_2, (N_2-1)a}\bigg). 
\eea

\noindent
For a 2d lattice fermion, we remain the anti-periodic boundary condition on the temporal direction ($x_1$). 
The fermion field satisfies the periodic boundary condition in another direction ($x_2$). 
\\

\noindent 
The topological charge only depends on a gauge configuration. 
If one uses plaquettes to define a topological charge, it can be non-trivial.  
Therefore, one cannot use the lattice artifact of fermion to imply that this lattice model only has a trivia sector.
Hence our proof only suggests that this lattice model necessarily combines other methods of defining a topological charge.

\section{Hybrid Monte Carlo Simulation}
\label{sec:4}
\begin{figure}[!htb]
\includegraphics[width=0.5\textwidth]{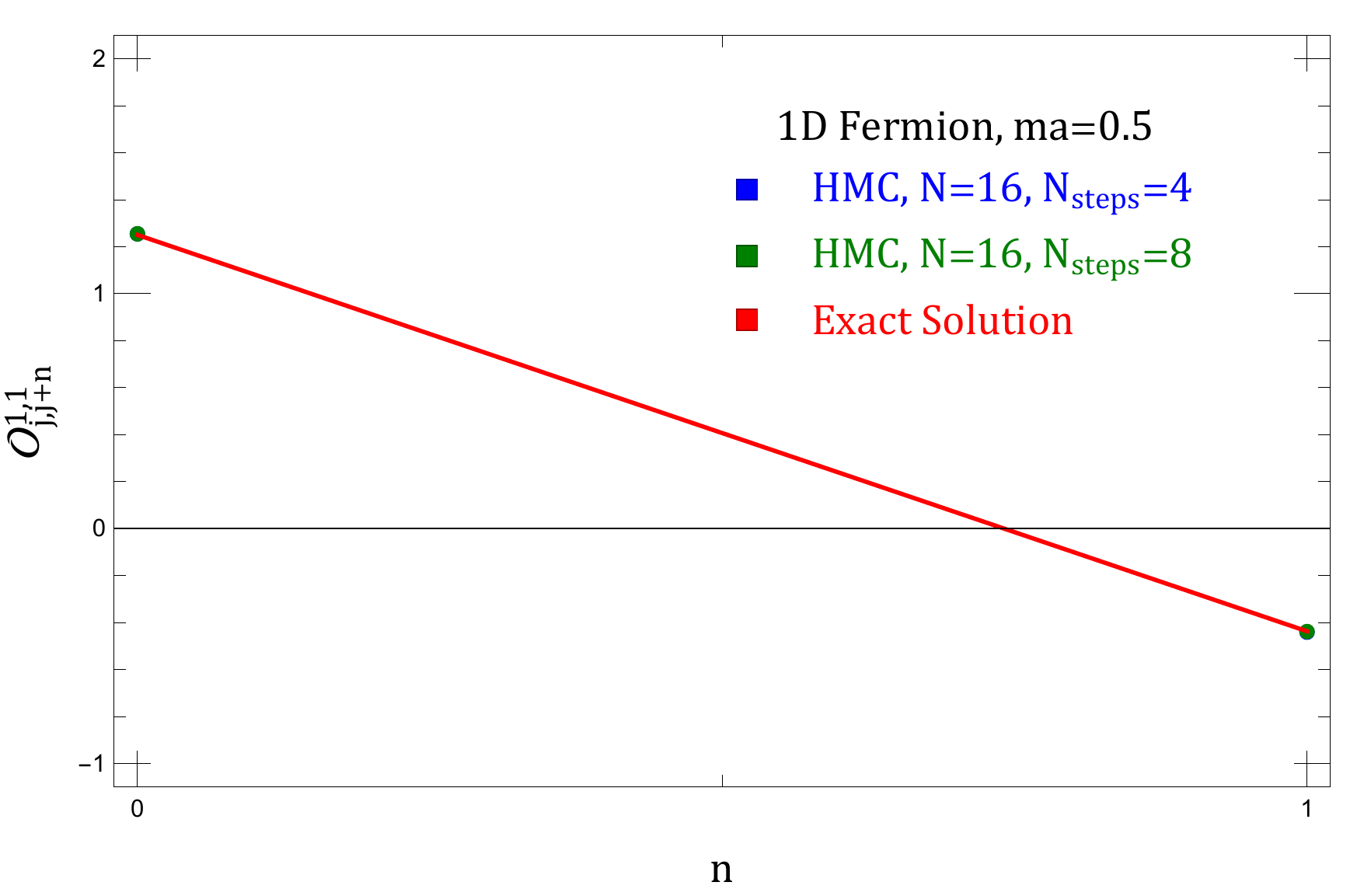}
\includegraphics[width=0.5\textwidth]{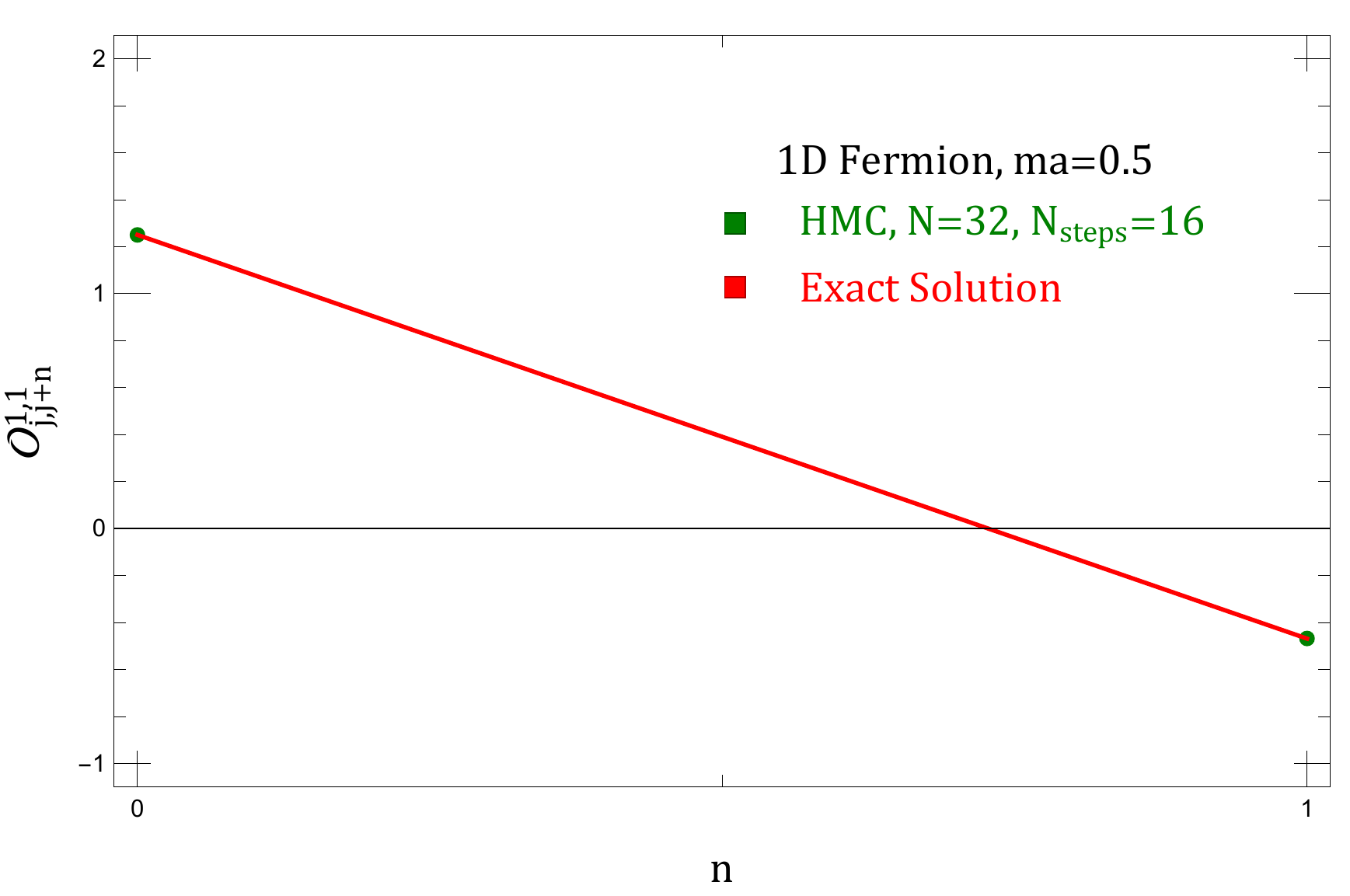}
\includegraphics[width=0.5\textwidth]{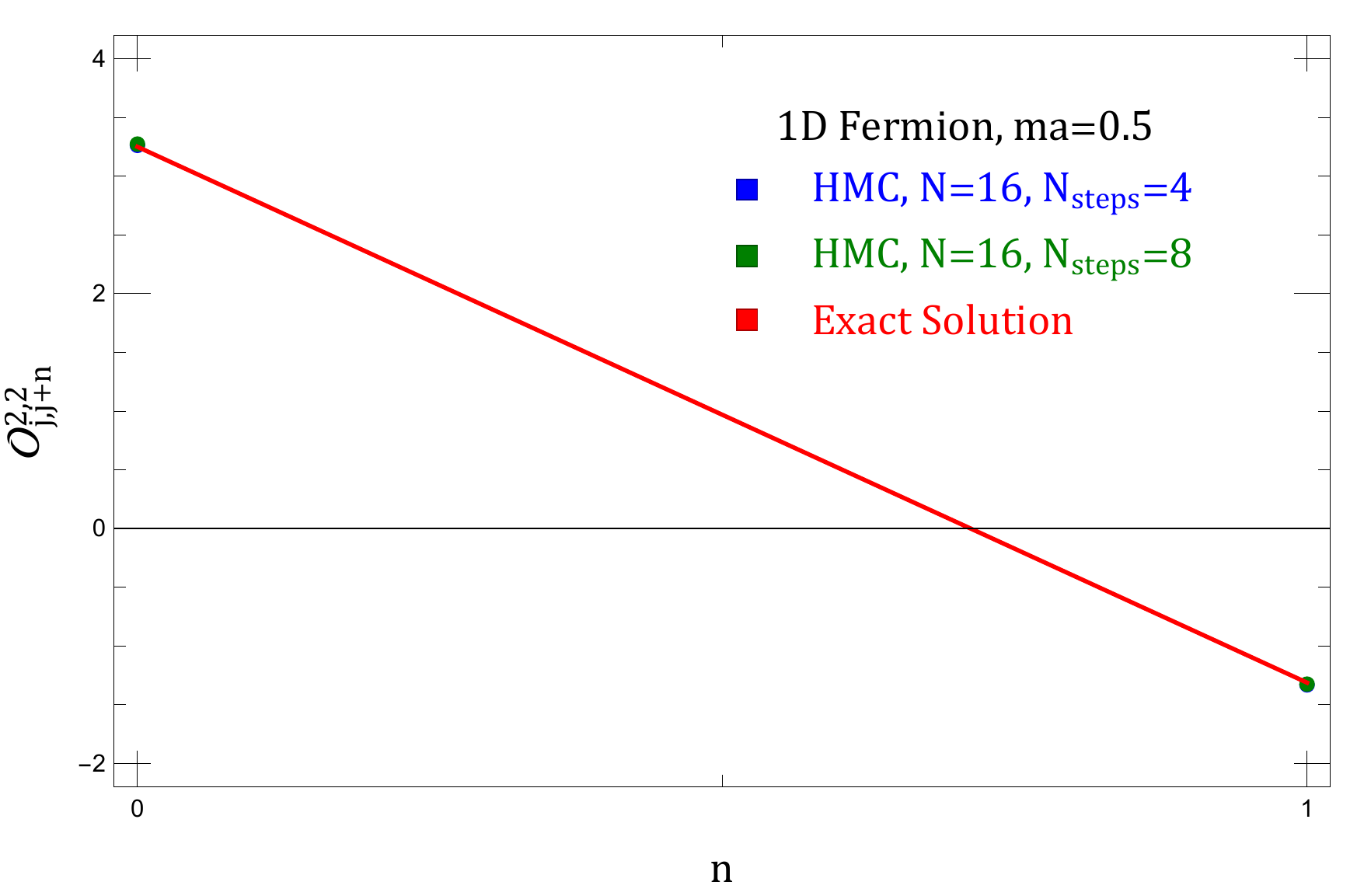}
\includegraphics[width=0.5\textwidth]{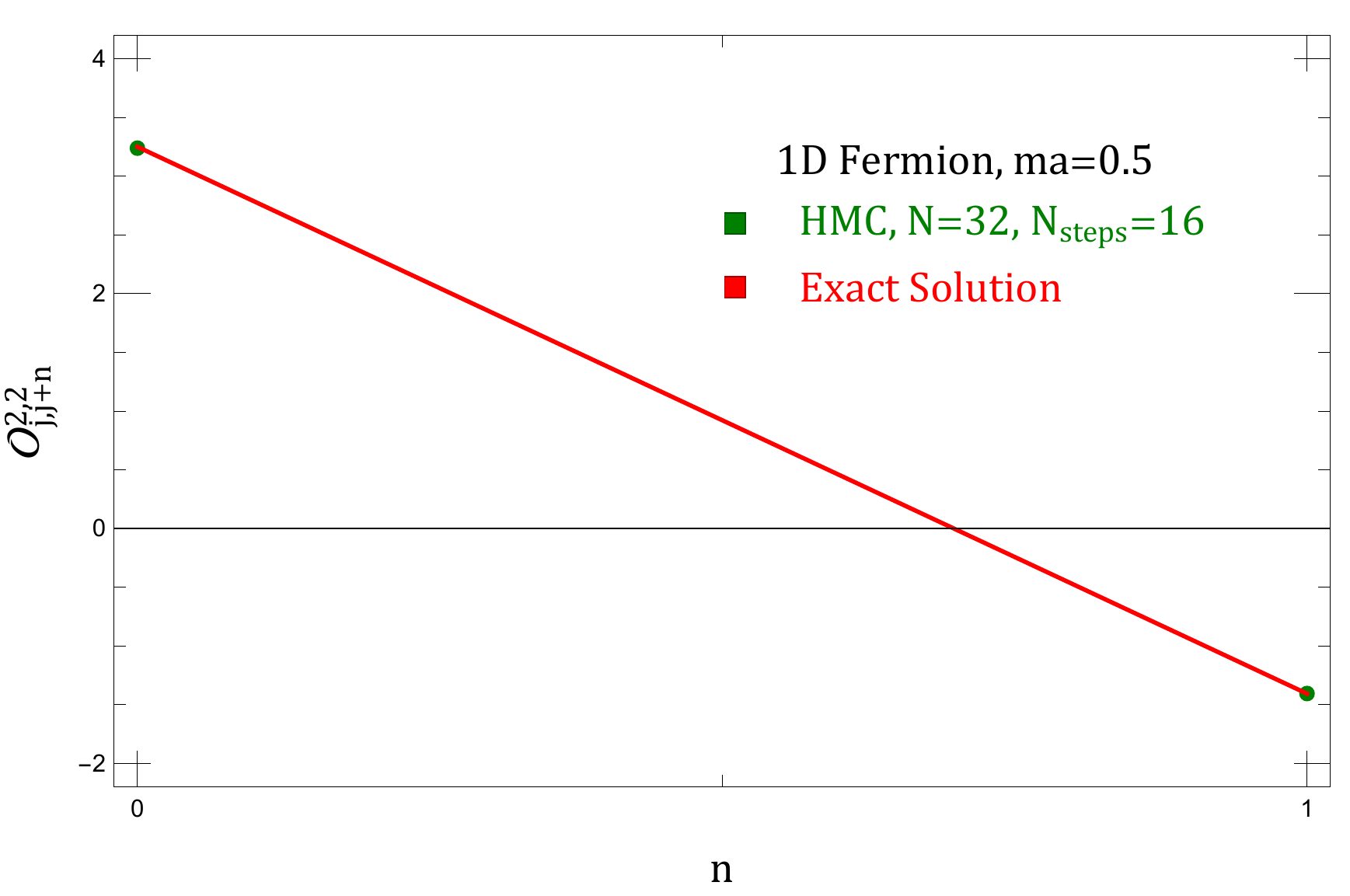}
\caption{We use the Hybrid Monte Carlo (HMC) to get the consistent result with the exact solution. 
The number of measurement is $2^{12}$ sweeps with thermalization $2^6$ sweeps and measure intervals $2^5$ sweeps. 
The error bars are less than $1\%$. 
The $N_{\mathrm{steps}}$ is the number of molecular dynamics steps.} 
\label{1st}
\end{figure}

\begin{figure}[!htb]
\includegraphics[width=0.5\textwidth]{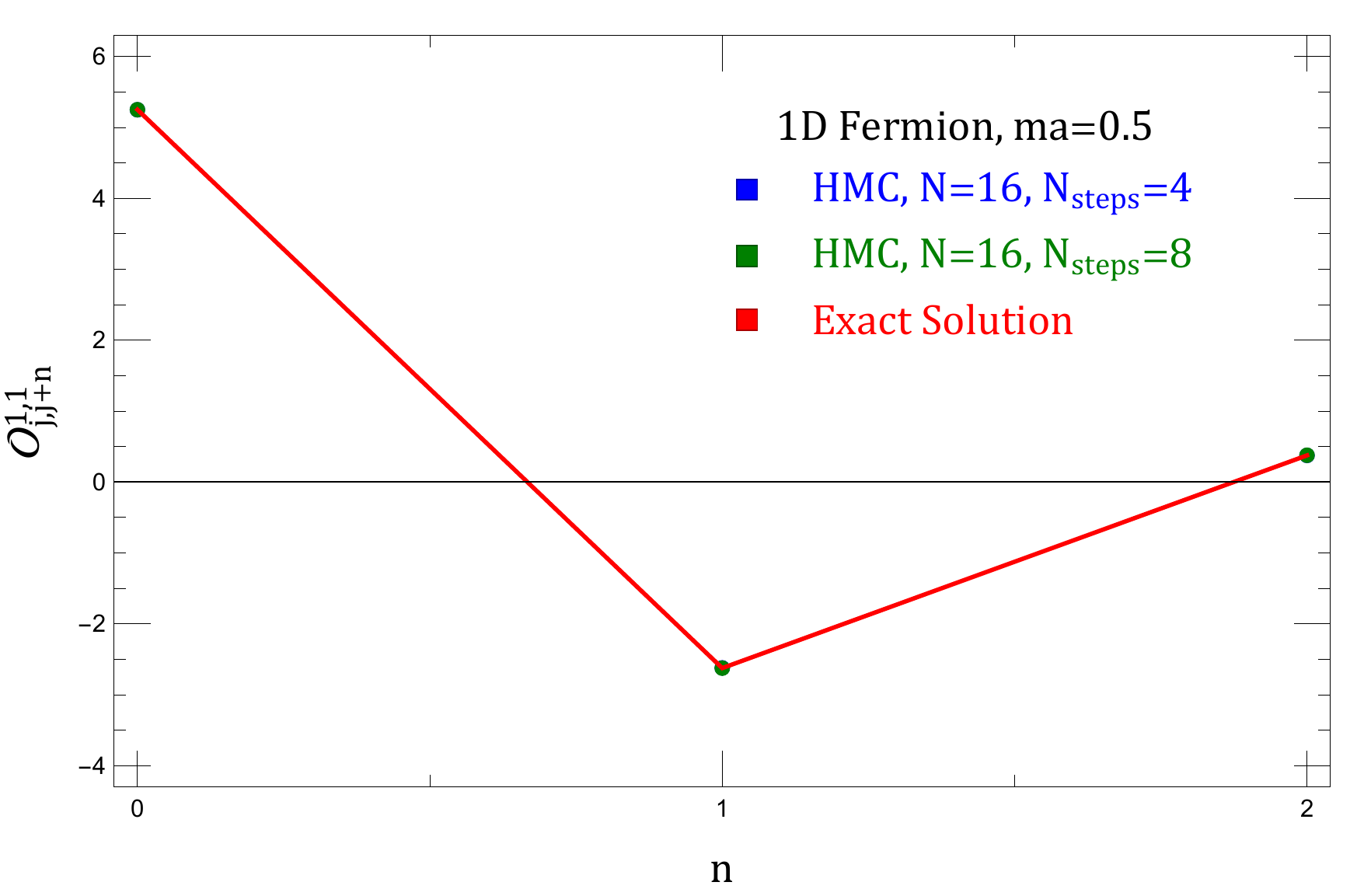}
\includegraphics[width=0.5\textwidth]{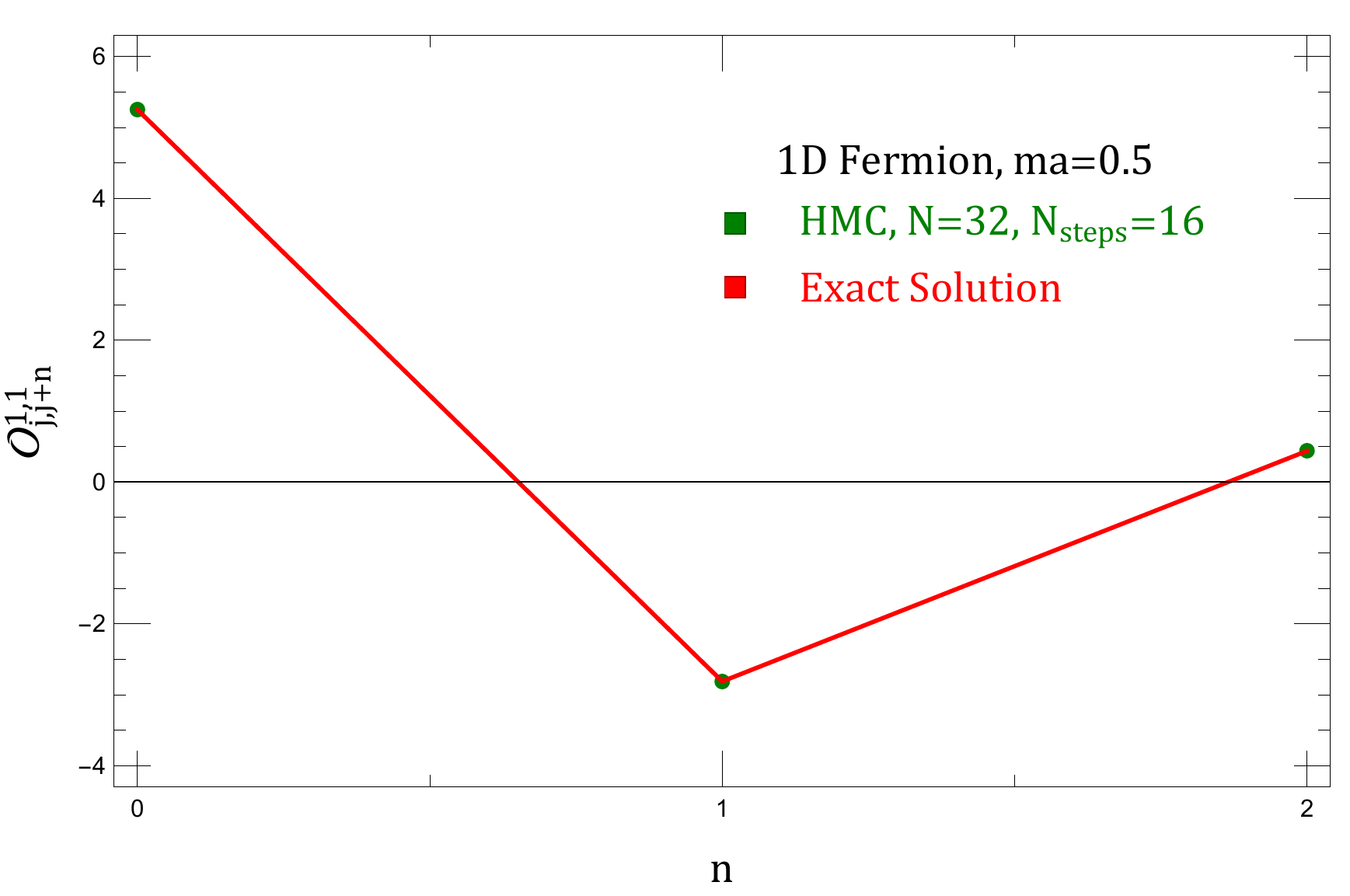}
\includegraphics[width=0.5\textwidth]{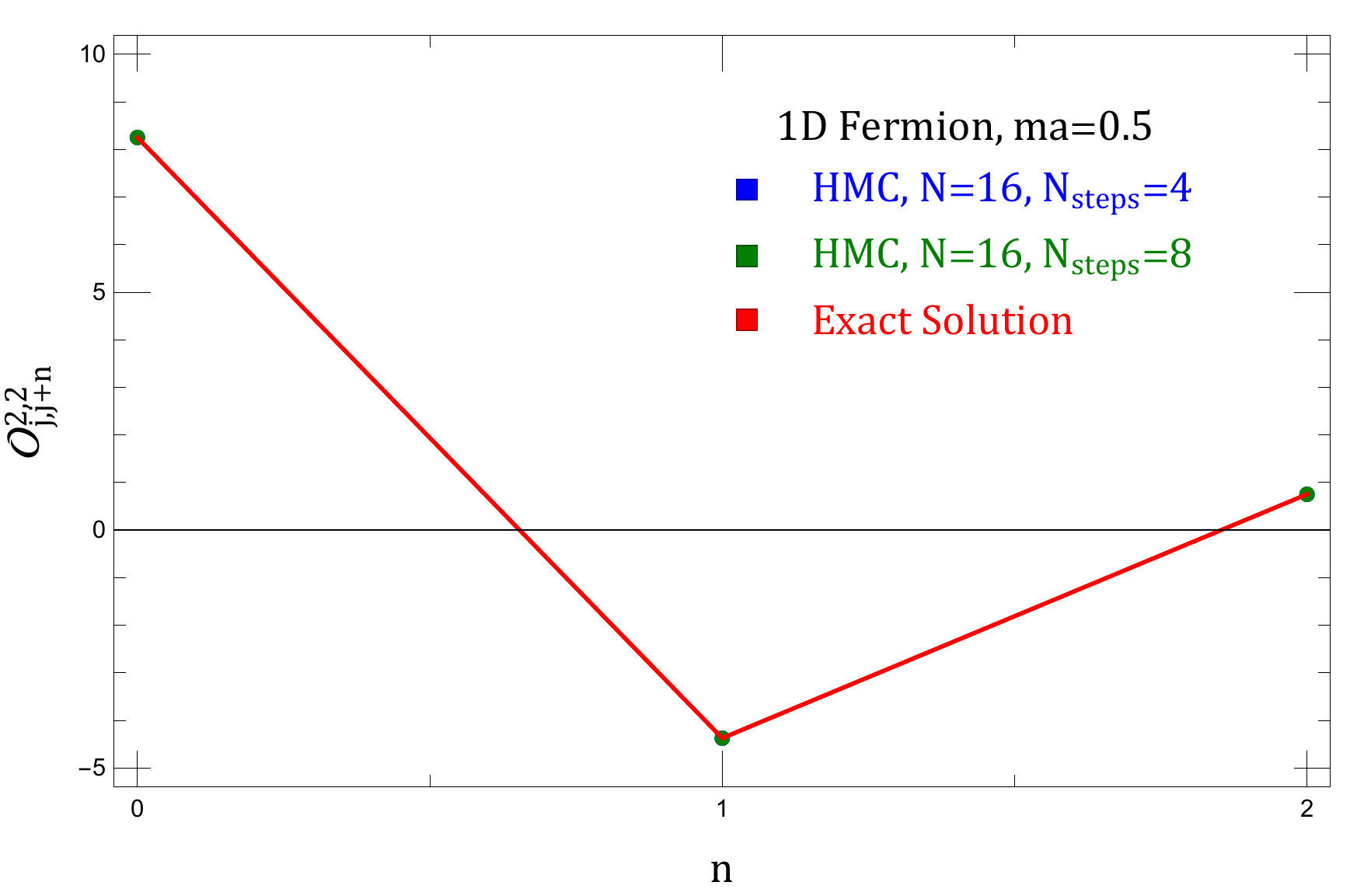}
\includegraphics[width=0.5\textwidth]{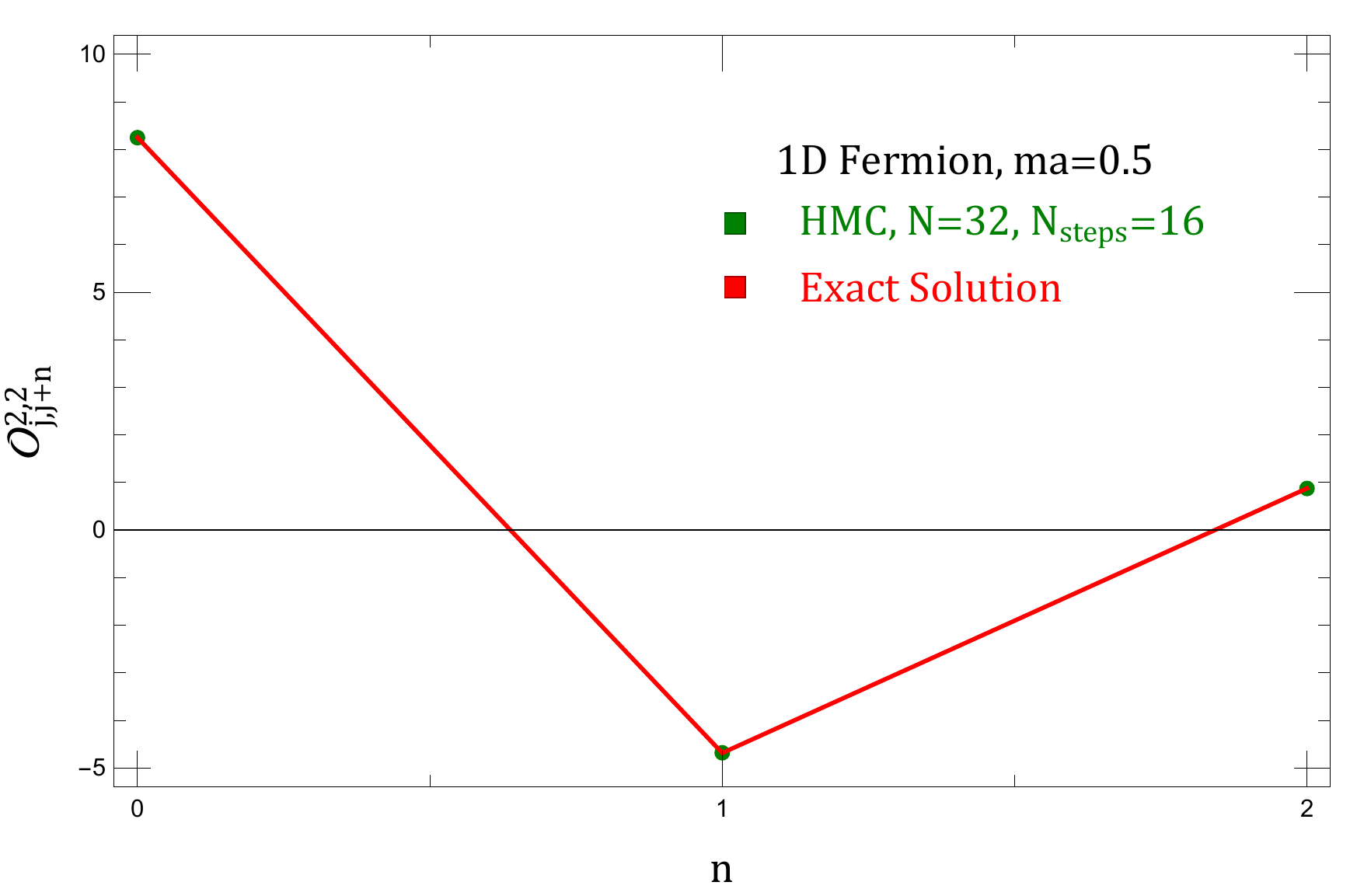}
\caption{We use the Hybrid Monte Carlo (HMC) to obtain the consistency from the exact solution. 
The number of measurement is $2^{18}$ sweeps with thermalization $2^7$ sweeps and measure intervals $2^6$ sweeps. 
The error bars are less than $1\%$. 
The $N_{\mathrm{steps}}$ is the number of molecular dynamics steps.
}
\label{2nd}
\end{figure}

\noindent 
One lattice model is necessary to show how practical a simulation is in the end. 
As in our discussion, the Dirac matrix of the forward finite-difference cannot recover the $\gamma_5$-hermiticity for non-physical modes. 
In other words, the Dirac matrix does not have a continuum limit in general. 
The Monte Carlo simulation, in general, relies on the determinant of a Dirac matrix for an importance sampling. 
Now we hope to fill the gap between the theoretical formulation and practical implementation. 
For two fermion fields in 1d with a degenerate mass, the lattice action is
\bea
S_{FD}=a\sum_{n=0}^{N-1}\bigg(\bar{\psi}_1(n)\big(D(n)+m\big)\psi_1(n)+
\bar{\psi}_2(n)\big(-D^{\dagger}(n)+m\big)\psi_2(n)\bigg).
\eea
Here we adopt the forward finite-difference with the first-order accuracy for the $\psi_1$. 
Adopting the backward finite-difference with the same accuracy is for the $\psi_2$.
After we integrate out the fermion fields, we obtain a non-negative determinant:
\bea
\det(D+m)\det(-D^{\dagger}+m)
&=&\det(D+m)\det\big(\gamma_5(-D^{\dagger}+m)\gamma_5\big)
\nn\\
&=&|\det(D+m)|^2. 
\eea 
We can introduce the pseudo-fermion field (bosonic field $\phi_f$) to rewrite the partition function as in the following
\bea
&&
\int {\cal D}\bar{\psi}{\cal D}\psi\ \exp(-S_{FD})
\nn\\
&\sim&
\int {\cal D}\phi_{f, R}{\cal D}\phi_{f, I}\ \exp\big(-\phi_f^{\dagger}\big((D+m)(D^{\dagger}+m)\big)^{-1}\phi_f\big),
\eea
where $\phi_f\equiv \phi_{f, R}+i\phi_{f, I}$. 
We then implement the Hybrid Monte Carlo algorithm to calculate: 
\bea
{\cal O}_{jk}^{\alpha\beta}\equiv
\frac{1}{2}
\langle \phi^{\dagger \alpha}_{f, j}\phi^{\beta}_{f, k}
+\phi^{\dagger \beta}_{f, k}\phi^{\alpha}_{f,j}
\rangle=\big((D+m)(D+m)^{\dagger}\big)_{jk}^{\alpha\beta}, 
\eea 
where $j, k=1, 2, \cdots, N$; $\alpha, \beta=1, 2$. 
We compare the exact solution to the numerical result for $N=$16 and 32 in Fig. \ref{1st}. 
We replace the first-order derivative with the second-order derivative and show the comparison in Fig. \ref{2nd}. 
The analysis shows that the numbers of thermalization and auto-correlation time are not high. 
Therefore, we expect that this lattice model can have a practical implementation.
The issue of non-physical modes should not give trouble to the Monte Carlo simulation. 
When including the interaction between fermions and gauge fields, it is necessary to average over $2^d$ possible orientations \cite{Stamatescu:1993ga, Sadooghi:1996ip}. 
For even flavor cases, one can apply our numerical algorithm to avoid the sign problem.

\section{Outlook}
\label{sec:5}
\noindent
We know that the transition of topological charge (defined by the index theorem \cite{Atiyah:1968mp, Atiyah:1970ws, Atiyah:1971rm}) is problematic in an overlap formulation. 
People still do not figure out the problem. 
The continuum limit in the lattice topological charge is subtle \cite{Chiu:2001bg, Chiu:2001ja}. 
As in our study, the Dirac matrix of a forward lattice formulation always shows a zero topological charge. 
The lattice chiral-symmetry cannot go back to the continuum symmetry for a non-physical mode under the continuum limit. 
Now one can extend our study to the 2d Schwinger model for solving this issue. 
The 2d theory has various ways to define a topological charge (like plaquette or index theorem). 
One can compare the result of continuum chiral symmetry to the lattice chiral symmetry case. 
Hence the issue due to the lattice artifact in topological charge or chiral symmetry should be clear. 
In 2d, one can also obtain all eigenvalues of a Dirac matrix without restricting to a low-lying mode. 
Hence the Schwinger model should be proper for a clean test before implementing a non-symmetrized finite-difference to LQCD. 

\section*{Acknowledgments}
\noindent 
Chen-Te Ma would like to thank Nan-Peng Ma for his encouragement.
\\

\noindent
Xingyu Guo acknowledges the Guangdong Major Project of Basic and Applied Basic Research No. 2020B0301030008 and NSFC Grant No.11905066.
Chen-Te Ma acknowledges the YST Program of the APCTP; 
China Postdoctoral Science Foundation, Postdoctoral General Funding: Second Class (Grant No. 2019M652926). 
Hui Zhang acknowledges the NSFC under Project No. 12047523. 


\end{document}